\documentclass[twocolumn,showpacs,preprintnumbers,amsmath,amssymb]{revtex4}

\usepackage{graphicx}
\usepackage{dcolumn}
\usepackage{bm}

\begin{document}

\preprint{APS/123-QED}

\title{Kink ratchets in the Klein-Gordon lattice free of the Peierls-Nabarro potential }

\author{Sergey V. Dmitriev,$^1$ Avinash Khare,$^2$ and Sergey V. Suchkov$^1$}
\affiliation{ $^1$Institute for Metals
Superplasticity Problems RAS, 450001 Ufa, Khalturina 39, Russia \\
$^2$Institute of Physics, Bhubaneswar, Orissa 751005, India
 }
\date{\today}

\begin{abstract}
A discrete Klein-Gordon model with asymmetric potential that
supports kinks free of the Peierls-Nabarro potential (PNp) is
constructed. Undamped ratchet of kinks under harmonic AC driving
force is investigated in this model numerically and contrasted
with the kink ratchets in the conventional discrete model where
kinks experience the PNp. We show that the PNp-free kinks exhibit
ratchet dynamics very much different from that reported for the
conventional lattice kinks which experience PNp
[see, e.g., Phys. Rev. E {\bf 73},
066621 (2006)]. Particularly, we could not observe any
significant influence of the discreteness parameter on the
acceleration of PNp-free kinks induced by the AC driving.
\end{abstract}


\maketitle

\section{Introduction} \label{sec:Introduction}

Ratchet dynamics of a point-like particle or a quasi-particle such
as soliton is the motion of the particle in a certain direction
under AC force whose average is zero. Ratchet transport phenomenon
can be observed under the following two conditions: (i) the system
must be out of thermal equilibrium and (ii) the space-time
symmetry of the system must be violated
\cite{BraunKivshar:2004,FYZ:2000,Reimann:2001,Reimann:2002}.
Ratchet dynamics has been receiving growing attention of
researchers from different fields ranging from biology
\cite{Cell:2002,Genetics:2008} and molecular motors
\cite{Motors,Motors1,Motors2,Motors3,Motors4}, through
superconducting Josephson juctions \cite{JJ,JJ1,JJ2,JJ3} and
nonlinear optics \cite{OptLett:2006}, to Bose-Einstein condensate
\cite{BEC:2008} and solid state physics \cite{SSP:2008}.

Soliton ratchets were first studied by Marchesoni
\cite{Marchesoni:1996} for the overdamped Klein-Gordon equation.
Unlike point-like particles, solitons can have internal
vibrational modes \cite{IntModes} and these modes can strongly
affect the ratchet dynamics especially for underdamped case
\cite{Modes1,Modes2}. So far, ratchet dynamics have been studied
for a number of continuum soliton bearing systems
\cite{Marchesoni:1996,Modes1,Modes2,Cont1,Cont2,Cont3,Cont4,Cont5,Cont6},
while in many applications ratchet was observed for discrete kinks
\cite{JJ1,JJ2,JJ3,OptLett:2006}. Effects of discreteness on kink
ratchets have been studied by Zolotaryuk and Salerno
\cite{Discrete1}. They have found that in comparison with the
continuum case, the discrete case shows a number of new features:
nonzero depinning threshold for the driving amplitude, locking to
the rational fractions of the driving frequency, and diffusive
ratchet motion in the case of weak intersite coupling. For the
damped, driven Frenkel-Kontorova chain, which is the discrete
analog of the sine-Gordon equation, Martinez and Chacon have shown
that phase disorder introduced into the asymmetric periodic
driving force can enhance the ratchet effect \cite{Discrete2}.

In discrete systems translational invariance is typically lost and
static solitary waves usually cannot be placed arbitrarily with
respect to the lattice but only in the positions corresponding to
the extremums of the Peirls-Nabarro potential (PNp), induced by
the lattice. Configurations corresponding to the maximums of PNp
are unstable while those corresponding to the minimums are stable.
It has been found that the presence of PNp makes kink ratchets
much more complicated in comparison to the continuum case
\cite{Discrete1}.

On the other hand, in the recent past, several different families
of discrete Klein-Gordon systems supporting translationally
invariant static solutions with arbitrary shift along the lattice
have been derived and investigated
\cite{PhysicaD,SpeightKleinGordon,BenderTovbis,JPA2005,Roy,CKMS_PRE2005,BOP,DKY_JPA_2006Mom,oxt1,DKYF_PRE2006,Coulomb,DKKS2007_BOP,ArXivKDS,SpeedyKinks,BarHeerden}.
Such discrete models are often called exceptional. Physical
properties of solitary waves in the exceptional discrete models
were found to be very much different from their conventional
counterparts. For instance, they can support conservation of
momentum \cite{PhysicaD,DKY_JPA_2006Mom} and can support kinks
which move with a special \cite{SpeedyKinks} or arbitrary
\cite{BarHeerden} velocity without emitting radiation.
Peculiarities of kink collisions in such models have been
investigated in the work \cite{Roy}. Static solutions in the
exceptional discrete systems possess the translational Goldstone
mode \cite{JPA2005,Roy}. This means that such static solutions are
not trapped by the lattice and can be accelerated by arbitrary
weak external field. Exceptional discrete models can describe
physically meaningful systems \cite{Coulomb}, that is why
investigation of physical properties of such systems is very
important.

In the present study we continue the investigation of physical
properties of the exceptional discrete models that support static
kinks free of PNp. The main goal of the study is to see how the
special properties of the kinks can influence their undamped
ratchet dynamics under single-harmonic driving.

In order to achieve this goal we need to construct an exceptional
discrete Klein-Gordon model with asymmetric background potential.
It is important that the constructed model be Hamiltonian,
otherwise energy increase in the system can be observed even for
small-amplitude driving with the frequency laying outside the
phonon band.

%
%

The paper is organized in five Sections. In Sec. \ref{sec:Setup},
we describe two discrete, Hamiltonian Klein-Gordon systems, with
and without PNp, having asymmetric on-site potential and having
common continuum limit. In Sec. \ref{sec:Kinks}, we compare the
properties of static kinks in these two models and then in Sec.
\ref{sec:Retchet} study kink ratchets mostly for the PNp-free
model. Section \ref{sec:Conclusions} concludes the paper.

\section{Discrete Klein-Gordon models with asymmetric potential}
\label{sec:Setup}

\subsection{General formulation}
\label{sec:SetupGeneral}

The Klein-Gordon field has the Hamiltonian
\begin{equation} \label{PotEn}
   H = \frac{1}{2} \int_{-\infty}^{\infty} \left[\phi_{t}^2 + \phi_{x}^2
   +2V(\phi) \right]dx\,,
\end{equation}
where $\phi(x,t)$ is the unknown field and $V(\phi)$ is a given
potential function. The corresponding equation of motion is
\begin{equation} \label{KG}
   \phi_{tt} = \phi_{xx} - V'(\phi)\,,
\end{equation}
where $V'(\phi)=dV/d\phi$.

Equation (\ref{KG}) will be discretized on the lattice $x=nh$,
where $n=0,\pm 1, \pm 2 ...$, and $h$ is the lattice spacing.
Traditional discretization of Eq. (\ref{KG}) is
\begin{equation} \label{Traditional}
   \ddot{\phi}_{n}= \frac{1}{h^2}(\phi_{n-1} -2\phi_{n} +\phi_{n+1})
   -V'(\phi_{n}).
\end{equation}
Using the discretized first integral (DFI) approach offered in
\cite{JPA2005} and developed in \cite{DKYF_PRE2006} one can
construct a discrete model whose static version is an integrable
map. Following this method we begin with the first integral of
static Eq. (\ref{KG}), $\phi_x^2 - 2V(\phi) +C = 0$, where $C$ is
the integration constant. The first integral can also be taken in
the following modified form \cite{JPA2005}
\begin{eqnarray} \label{FirstIntv}
   v(x) \equiv \phi_x - \sqrt{2V(\phi)-C} = 0 \,.
\end{eqnarray}
Next step is to rewrite the Hamiltonian Eq. (\ref{PotEn}) in terms
of $v(x)$ as follows
\begin{equation} \label{PotEnFI}
   H =\frac{1}{2} \int_{-\infty}^{\infty} \Big\{ \phi_{t}^2 +
   \left[v(x)\right]^2 +2\phi_x\sqrt{2V(\phi)-C}\Big\}dx\,,
\end{equation}
where we omitted the constant term.

The first integral Eq. (\ref{FirstIntv}) can be discretized as
follows
\begin{eqnarray} \label{TwoPointMapv}
   \tilde{v}(\phi_{n-1},\phi_{n}) \equiv \frac{\phi_{n}
   -\phi_{n-1}}{h} 
   -\sqrt{2V(\phi_{n-1},\phi_{n}) - C} = 0,
\end{eqnarray}
where we demand that $V(\phi_{n-1},\phi_{n}) \rightarrow V(\phi)$
in the continuum limit ($h \rightarrow 0$). Thus we obtain the
discrete version of the Hamiltonian Eq. (\ref{PotEn})
\begin{eqnarray} \label{HamiltonianDiscr}
   H = \frac{1}{2}\sum_{n} \Big\{ \dot{\phi}_n^2
   +\left[\tilde{v}(\phi_{n-1},
   \phi_n)\right]^2 \nonumber \\
   +2\frac{\phi_n-\phi_{n-1}}{h}\sqrt{2V(\phi_{n-1},\phi_n)-C}\Big\}\,.
\end{eqnarray}
Final step is to discretize the background potential as suggested
by Speight \cite{SpeightKleinGordon},
\begin{eqnarray} \label{PNpFreeCondition}
   \sqrt {2V\left( {\phi_{n-1},\phi_n} \right) -C}  =
   \frac{{G(\phi_n) - G(\phi_{n-1})}}{{\phi_n - \phi_{n-1}}}, \nonumber \\
   {\rm where} \quad G'\left( \phi  \right) = \sqrt {2V(\phi)-C}.
\end{eqnarray}
With this choice the last term of the Hamiltonian Eq.
(\ref{HamiltonianDiscr}) reduces to $(2/h)[G(\phi_n) -
G(\phi_{n-1})]$ and it disappears in the telescopic summation.
Further, according to Eq. (\ref{PNpFreeCondition}), the
discretized first integral Eq. (\ref{TwoPointMapv}) assumes the
form
\begin{eqnarray} \label{TwoPointMapv1}
   v(\phi_{n-1},\phi_{n}) = \frac{\phi_{n}
   -\phi_{n-1}}{h}  
   -\frac{G(\phi_n) - G(\phi_{n-1})}{\phi_n - \phi_{n-1}},
\end{eqnarray}
and the equations of motion derived from Eq.
(\ref{HamiltonianDiscr}) with $\tilde{v}(\phi_{n-1}, \phi_n) =
v(\phi_{n-1}, \phi_n)$ are
\begin{eqnarray} \label{HamiltPNpFree}
  \ddot{\phi}_n=-v(\phi_{n-1},\phi_n)\frac{\partial }{{\partial \phi_n}}
  v(\phi_{n-1},\phi_n) \nonumber \\
  -v(\phi_{n},\phi_{n+1})\frac{\partial }{{\partial \phi_n}}
  v(\phi_{n},\phi_{n+1}).
\end{eqnarray}
Obviously, equilibrium static solutions of this model can be found
from the two-point nonlinear map $v(\phi_{n-1},\phi_{n}) = 0$,
where $v(\phi_{n-1},\phi_{n})$ is given by Eq.
(\ref{TwoPointMapv1}). Such solutions can be constructed
iteratively starting from any admissible initial value
$\phi_{n-1}$ or $\phi_{n}$, and thus, the PNp is absent for such
family of equilibrium solutions.

\subsection{Polynomial asymmetric potential}
\label{sec:SetupPolynomial}

Klein-Gordon kink ratchets are possible if the on-site potential
or the driving force or both are asymmetric. We study the kink
ratchets under single-harmonic AC driving and thus, the on-site
potential must be asymmetric.

We take $G'$ in Eq. (\ref{PNpFreeCondition}) in the form of the
quartic polynomial function
\begin{eqnarray} \label{Gprime}
  G'(\phi) = a\phi^4 + b\phi^2 + c\phi + d\,,
\end{eqnarray}
where the cubic term was not taken into account because it can
always be removed by a proper shift $\phi \rightarrow \phi -
\phi_0$. Then the on-site potential (with $C=0$) is
\begin{eqnarray} \label{Vpoly}
  V(\phi)=\frac{1}{2}\left(a\phi^4 + b\phi^2 + c\phi + d \right)^2.
\end{eqnarray}
The simplest discrete Klein-Gordon model corresponding to this
potential (will be referred to as DKGM1), according to Eq.
(\ref{Traditional}), is
\begin{eqnarray} \label{TraditionalPoly}
  \ddot \phi_n  = \frac{1}{h^2}\left(\phi_{n - 1} - 2\phi_n
  + \phi_{n + 1} \right) \nonumber \\
  - \left(a\phi_n^4 + b\phi_n^2  +
  c\phi _n  + d \right)\left( 4a\phi_n^3 + 2b\phi_n + c
  \right),
\end{eqnarray}
and its Hamiltonian is
\begin{eqnarray} \label{HamiltDKGM1}
   H_1 = \frac{1}{2}\sum_{n} \Big\{ \dot{\phi}_n^2
   + \frac{1}{h^2}(\phi_n-\phi_{n-1})^2 \nonumber \\
   + \left(a\phi_n^4 + b\phi_n^2 + c\phi_n + d \right)^2 \Big\}\,.
\end{eqnarray}

 A more sophisticated discrete Klein-Gordon model (will be
referred to as DKGM2) is defined by Eq. (\ref{HamiltPNpFree}) with
\begin{eqnarray} \label{DFIPoly}
 v\left(\phi_{n - 1} ,\phi_n \right)
  = \frac{\phi _n  - \phi _{n - 1}}{h} \nonumber \\
  - \frac{a}{5}\left(\phi_n^4+ \phi_n^3 \phi_{n - 1}+
  \phi_n^2 \phi_{n - 1}^2+ \phi_n \phi _{n - 1}^3
  + \phi_{n - 1}^4 \right) \nonumber \\
  - \frac{b}{3}\left( {\phi _n^2  + \phi _n \phi _{n - 1}
  + \phi _{n - 1}^2 } \right) 
  - \frac{c}{2}\left( {\phi _n  + \phi _{n - 1} } \right) - d,
\end{eqnarray}
which was found by substituting Eq. (\ref{Gprime}) after
integrating into Eq. (\ref{TwoPointMapv1}). This model has the
Hamiltonian
\begin{eqnarray} \label{HamiltDKGM2}
   H_2 = \frac{1}{2}\sum_{n} \Big\{ \dot{\phi}_n^2
   + \left[v(\phi_{n-1},
   \phi_n)\right]^2 \Big\}\,.
\end{eqnarray}

\begin{figure}
\includegraphics{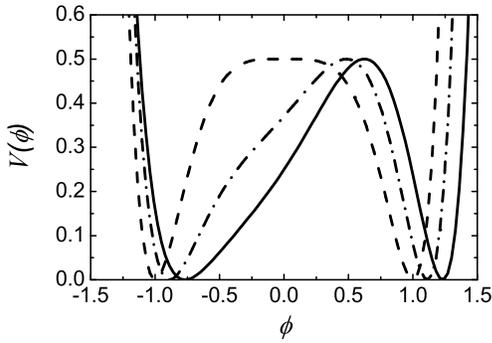}
\caption{On-site potential Eq. (\ref{Vpoly}) for $b=0$ and $a=-1$,
$c=0$, $d=1$ (dashed line); $a=-0.83988$, $c=0.382925$,
$d=0.860756$ (dash-dotted line); $a=-0.643049$, $c=0.626843$,
$d=0.706344$ (solid line).} \label{fig1}
\end{figure}

The asymmetry of the potential Eq. (\ref{Vpoly}) is controlled by
the parameter $c$ and for $c=0$ the potential is symmetric. The
on-site potential Eq. (\ref{Vpoly}) has four parameters. Let us
fix the height of the potential barrier equal to 0.5, the distance
between the two minima equal to 2. The asymmetry can be chosen by
setting a value of $c$. There is still one free parameter and we
will take $b=0$.

In Fig. \ref{fig1} we plot the on-site potential Eq. (\ref{Vpoly})
for $b=0$ and $a=-1$, $c=0$, $d=1$ (dashed line); $a=-0.83988$,
$c=0.382925$, $d=0.860756$ (dash-dotted line); $a=-0.643049$,
$c=0.626843$, $d=0.706344$ (solid line). One can see that the
asymmetry of the potential increases with $c$.

The asymmetric potential supports two vacuum solutions,
$\phi_n=\phi_1$ and $\phi_n=\phi_2$, where $\phi_1$ and $\phi_2$
are the coordinates of the two minima of the on-site potential.
Small-amplitude phonon vibrations of the form $\phi_n \sim
\exp[i(qn-\omega t)]$, where $q$ is the phonon wavenumber and
$\omega$ is the phonon frequency, have different spectra for
different vacuums.

Borders of the phonon bands for each vacuum can be found for DKGM2
from

\begin{eqnarray} \label{PhononBands1}
 \omega _1^2  = 28a^2 \phi _i^6  + 30ab\phi _i^4  + 20ac\phi _i^3
 \nonumber \\
+ 3\left( {2b^2  + 4ad} \right)\phi _i^2  + 6bc\phi _i  + c^2  +
2bd ,
\end{eqnarray}
\begin{eqnarray} \label{PhononBands2}
 \omega _2^2  = \frac{{28}}{{25}}a^2 \phi _i^6  + 2ab\phi _i^4  +
\frac{4}{5}ac\phi _i^3 \nonumber \\  + 6\left( {\frac{{b^2 }}{9} +
\frac{2}{5}ad} \right)\phi _i^2  + \frac{2}{3}bc\phi _i  +
\frac{2}{3}bd + \frac{4}{{h^2 }},
\end{eqnarray}
where $j=1,2$ and $\omega_1$ ($\omega_2$) corresponds to $q=0$
($q=\pi$).

For DKGM1 the borders corresponding to $q=0$ coincide with that
for DKGM2, while the borders corresponding to $q=\pi$ are
\begin{eqnarray} \label{PhononBands3}
\omega _2^2  = 28a^2 \phi _i^6  + 30ab\phi _i^4  + 20ac\phi _i^3 + \nonumber \\
6\left( {b^2  + 2ad} \right)\phi _i^2 + 6bc\phi _i + c^2  + 2bd +
\frac{4}{{h^2 }}.
\end{eqnarray}

Numerical results in this work will be obtained for the on-site
potential with the parameters $a=-0.643049$, $b=0$, $c=0.626843$,
$d=0.706344$ (shown by the solid line in Fig. \ref{fig1}). The
potential has minima at $\phi_1=-0.768678$ and $\phi_2=1.231321$
and a maximum at $\phi_{\max}=0.62462$.

\section{Properties of static kinks in two lattices} \label{sec:Kinks}

Before we proceed with the we need to study the properties of the
kinks in DKGM1 and DKGM2 because they will help us to interpret
the results of the kink ratchet dynamics study.

Equilibrium static kink solutions for the DKGM1 can be found
numerically while for DKGM2 they can be found iteratively using
Eq. (\ref{DFIPoly}) for any initial value of $\phi_n$ (or
$\phi_{n-1}$) lying between two minima of the on-site potential.

\begin{figure}
\includegraphics{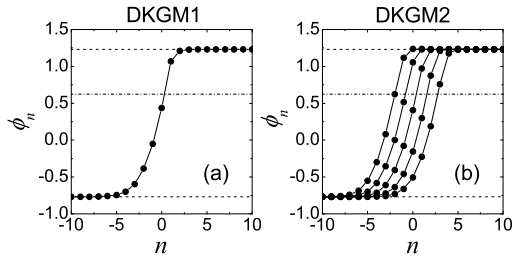}
\caption{(a) Stable equilibrium kink in DKGM1 and (b) a family of
equilibrium kinks in DKGM2. In both cases $h=0.6$. Dashed lines
show the locations of minima of the on-site potential and the
dash-dotted line the location of the maximum.} \label{fig2}
\end{figure}

In the DKGM1 there exists only one stable static kink
configuration [shown in Fig. \ref{fig2} (a)], corresponding to the
minimum of the PNp. Static kinks in the DKGM2 do not experience
the PNp and they can be placed anywhere with respect to the
lattice. A family of equilibrium kinks is presented in Fig.
\ref{fig2} (b). Kinks in both models have asymmetric shape because
of the asymmetry of the on-site potential.

\begin{figure}
\includegraphics{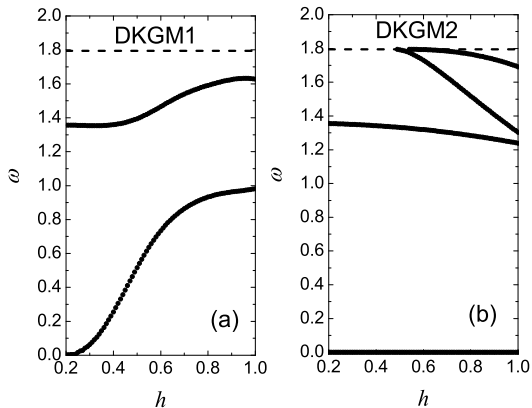}
\caption{Kink's internal mode frequencies as the functions of the
discreteness parameter for (a) DKGM1 and (b) DKGM2. Dashed line
shows the bottom edge of the phonon band. Model parameters (here
and in the following): $a=-0.643049$, $b=0$, $c=0.626843$,
$d=0.706344$.} \label{fig3}
\end{figure}


Small-amplitude oscillation spectra for the chains containing a
kink are presented in Fig. \ref{fig3} for different values of the
discreteness parameter $h$. Dashed horizontal line shows the lower
edge of the phonon band which is the same for both DKGM1 and DKGM2
and can be found from Eq. (\ref{PhononBands1}) for $j=1$ (soft
minimum). Presented spectra show the kink's internal vibrational
modes. Since both discrete models share the same continuum limit
[defined by Eq. (\ref{KG}) and Eq. (\ref{Vpoly})], for small
discreteness ($h<0.25$) their spectra are close. Kink in DKGM1
[see in (a)] possesses two internal modes, one of them is the
destroyed translational mode (for small $h$ it approaches zero
frequency). Kink in DKGM2 [see in (b)] possesses the
zero-frequency translational mode for any $h$, and for $h>0.48$
two new internal modes appear. Note that the spectrum in (b) was
calculated for the kink having a particle at the maximum of the
on-site potential and the kink internal mode frequencies within
the studied range of $h$ are only slightly dependent on the
location of the kink with respect to the lattice. For example,
maximal difference between the internal mode frequencies for
different kink positions with respect to the lattice observed at
$h=1$ is 0.9\%.


\section{Kink ratchets} \label{sec:Retchet}

To study kink ratchets we add to the right-hand sides of equations
of motion Eq. (\ref{TraditionalPoly}) (DKGM1) and Eq.
(\ref{HamiltPNpFree}) (DKGM2) the harmonic external force
\begin{eqnarray} \label{Rforce}
 F(t) = A\cos(\Omega t + \varphi),
\end{eqnarray}
with the amplitude $A$, frequency $\Omega$, and initial phase
$\varphi$.

The initial conditions are thus as follows: we have a static
equilibrium kink and at $t=0$ the force Eq. (\ref{Rforce}) is
turned "on".

\begin{figure}
\includegraphics{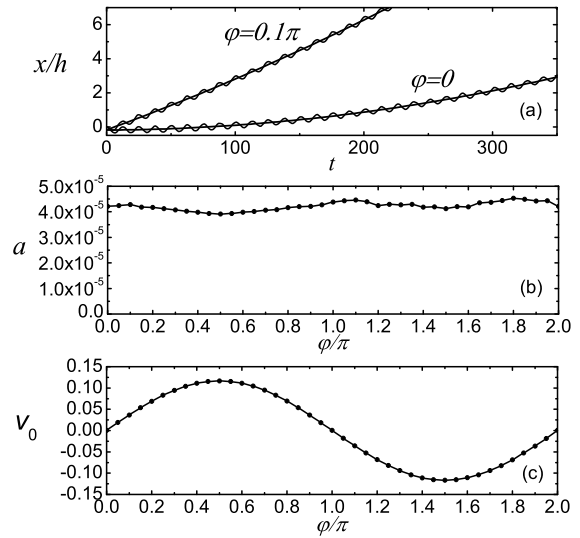}
\caption{(a) Kink motion in DKGM2 at $h=0.6$ for two different
values of initial phase of driving force $\varphi$ at the same
amplitude $A=0.04$ and frequency $\Omega=0.5$. Oscillating lines,
presenting the kink coordinate as the functions of time, are
fitted by the square parabolas. (b) Acceleration of the kink and
(c) initial velocity of the kink as the functions of $\varphi$.}
\label{fig5}
\end{figure}

\begin{figure}
\includegraphics{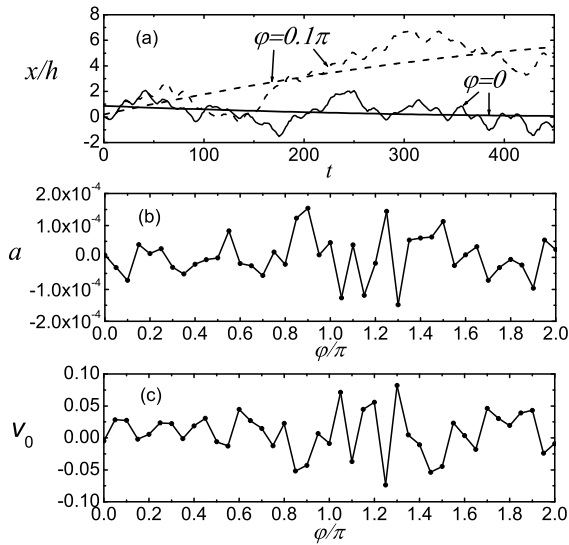}
\caption{Same as in Fig. \ref{fig5} but for DKGM1, where kink
experiences PNp.} \label{fig6}
\end{figure}

In contrast to the majority of studies on the soliton ratchets we
study the kink ratchets in the models that include no viscosity
terms. We thus restrict ourselves to the case of driving force
with a small amplitude ($A\le 0.04$) and with frequency lying
outside the phonon band (more precisely, below the phonon
spectrum), otherwise phonon modes will be excited and the analysis
of kink ratchets will become more complicated. Study of the
undamped ratchet makes it possible to directly measure the force
driving the kink.

For the models with damping terms it is customary to measure the
efficiency of ratchet by the averaged velocity of steady motion of
the soliton. This is not applicable to our case and instead, we
measure the acceleration of the kink, $a$. As it will be shown,
this approach works well for DKGM2, which is the primary subject
of present study. This is demonstrated by the numerical results
presented in Fig. \ref{fig5} where in (a) we show two examples of
kink's trajectories (lines oscillating with the frequency of
driving force, $\Omega=0.5$) and the least-square fit to these
lines by square parabola
\begin{eqnarray} \label{Parabola}
   x(t) = at^2 + v_0 t + x_0,
\end{eqnarray}
where $a$ is the net acceleration of the kink and $v_0$, $x_0$ are
the kink's initial velocity and coordinate, respectively. One can
see that kink's trajectories are fitted very well by the square
parabola, meaning that the motion of kink is uniformly accelerated
within the studied time domain.

The two trajectories shown in Fig. 5(a) correspond to different
initial phases $\varphi$ of the driving force Eq. (\ref{Rforce}),
while all other parameters are same: $A=0.04$, $\Omega=0.5$,
$h=0.6$. In the case of $\varphi=0$ kink does not get initial
momentum (in this case $v_0 = 0$) while in the case of
$\varphi=0.1\pi$ it does ($v_0 \neq 0$). However, acceleration $a$
of the kink in both cases is nearly same. In the panels (b) and
(c) of Fig. 5 we plot the acceleration $a$ and the initial
velocity $v_0$ of the kink, respectively, as the functions of the
initial phase of driving force, $\varphi$. It can be seen that
$v_0$ changes noticeably with $\varphi$ but its average over
$\varphi$ is zero. On the other hand, the acceleration of the kink
is practically independent of $\varphi$ and in the rest of the
paper we set $\varphi=0$.

We have also checked how the kink's acceleration $a$ depends on
the initial position of the static kink with respect to the
lattice, $x_0$, and found that $a$ practically does not depend on
$x_0$ for the chains with $h=0.3$, $h=0.6$, and $h=0.9$. Even for
the largest studied value of $h=0.9$ the difference between $a$
measured for kinks with different $x_0$ was within the numerical
error.

In Fig. \ref{fig6} we show same as in Fig. \ref{fig5} but for
DKGM1 where kink experiences PNp. The results in this case are
strikingly different. Kink's trajectories are now irregular and
their least-square fit by square parabola does not make much
sense. Nevertheless, in the panels (b) and (c) we present the
values of $a$ and $v_0$ obtained from such fit of trajectories
corresponding to various $\varphi$. Both $a$ and $v_0$ vary
irregularly. We thus conclude that presence of PNp largely affects
the ratchet dynamics of kink in our settings. In contrast to the
DKGM2, where PNp is absent, motion of kinks in DKGM1 is not
uniformly accelerated, at least for the range of parameters
studied in this work, i.e., for rather small amplitude $A$ of the
driving force. This is true already at moderate degree of
discreteness, $h=0.6$, and the influence of PNp increases with
increase in $h$.

Now we turn back to the DKGM2 and study the influence of the
driving force parameters $A$ and $\Omega$ and the discreteness
parameter $h$ on the uniformly accelerated dynamics of the kink.

\begin{figure}
\includegraphics{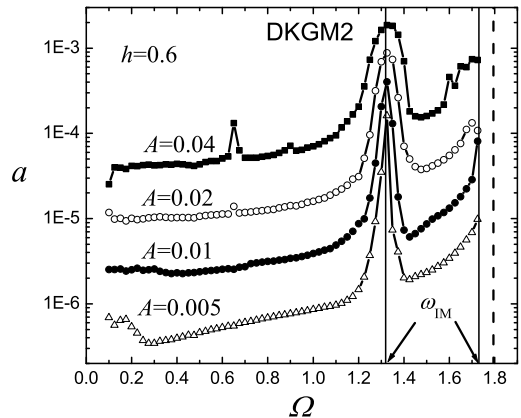}
\caption{Acceleration of kink in DKGM2 as the function of
frequency of driving force for different values of the force
amplitude, $A$, as specified near each curve. Initial phase of the
force $\varphi=0$ and the discreteness parameter $h=0.6$. Vertical
solid lines show the frequencies of the kink's internal modes and
the vertical dashed line shows the lower edge of the phonon band.}
\label{fig7}
\end{figure}

\begin{figure}
\includegraphics{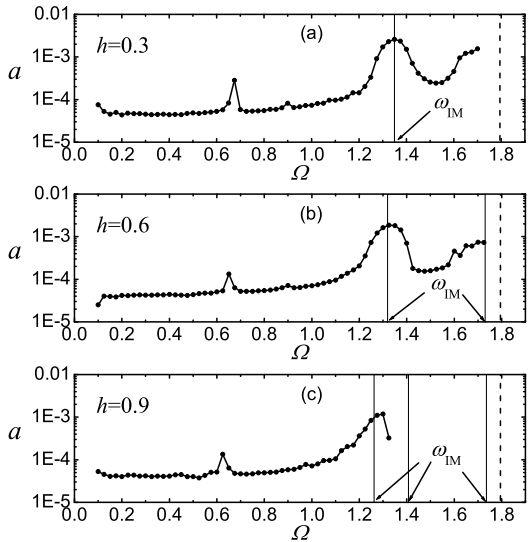}
\caption{Acceleration of kink in DKGM2 as the function of
frequency of driving force for different values of the
discreteness parameter $h$, as specified in each panel. Amplitude
of the ac driving force is $A=0.04$ and initial phase of the force
is $\varphi=0$. Vertical solid lines show the frequencies of the
kink's internal modes and the vertical dashed line shows the lower
edge of the phonon band.} \label{fig8}
\end{figure}

Results presented in Fig. \ref{fig7} were obtained for $h=0.6$.
Here we plot how the kink's acceleration $a$ depends on the
driving force frequency $\Omega$ at different values of the
amplitude of the force $A$, as indicated near each curve. Vertical
solid lines show the frequencies of the kink's internal modes and
the vertical dashed line shows the lower edge of the phonon band.
It is readily seen that the acceleration of the kink increases by
one or even two orders of magnitude (note the logarithmic scale
for the ordinate) when the driving force frequency $\Omega$
approaches the frequency of kink's internal mode $\omega_{\rm
IM}=1.32$ (also note a smaller peak at $\omega_{\rm IM}/2$). This
finding agrees well with earlier observations on the role of the
kink's internal modes on ratchet dynamics in the underdamped case
\cite{BraunKivshar:2004,Modes1,Cont5,Cont6,Discrete1}. One can
also see that $a$ increases when $\Omega$ approaches kink's
another internal mode frequency $\omega_{\rm IM}=1.73$. However,
this increase can also be attributed to the fact that the phonon
band edge ($\omega_1=1.795$) is also approached. A special
investigation is required to clarify which of these two factors
plays major role in increasing $a$. Looking at Fig. \ref{fig7},
one can also notice that the increase of the driving force
amplitude $A$ by one order of magnitude has resulted in the
increase in $a$ by two orders of magnitude and thus, $a \sim A^2$.
Note that the scaling rule $\left\langle v \right\rangle \sim
A^2$, where $\left\langle v \right\rangle$ is the averaged
stationary kink velocity, has been reported for the soliton
ratchets \cite{BraunKivshar:2004}.

Finally, we discuss the influence of the discreteness parameter
$h$ on the kink's acceleration $a$ at various driving force
frequencies $\Omega$ (see Fig. \ref{fig8}). Here we set $A=0.04$
and consider the cases of relatively weak ($h=0.3$), moderate
($h=0.6$), and strong ($h=0.9$) discreteness. Again, vertical
solid lines show the frequencies of the kink's internal modes and
the vertical dashed lines show the lower edge of the phonon band.
Remarkably, for $\Omega<1.2$ the results are very close for all
three values of $h$. This can be interpreted in such a way that
for the discrete model supporting PNp-free kinks the ratchet
dynamics is more like in the continuum case. The difference in the
results that appears for $\Omega>1.2$ is related to the kink's
internal modes, whose frequencies are $h$-dependent [see Fig.
\ref{fig3}(b)]. We note that there is no numerical data in Fig.
\ref{fig8}(c) for $\Omega>1.325$. Above this frequency there is a
mixed influence of the two kink's internal modes and motion of the
kink becomes different from uniformly accelerated so that one
cannot assign any particular value of $a$ to it.

\section{Conclusions} \label{sec:Conclusions}

Undamped ratchet dynamics of discrete Klein-Gordon kinks free of
the Peirls-Nabarro potential was investigated numerically. For
this purpose a lattice with asymmetric on-site potential was
constructed.

It was found that, typically, in the presence of single-harmonic
AC driving and in the absence of damping, PNp-free kink dynamics
is uniformly accelerated until its velocity becomes too large and
radiation losses start to contribute to the dynamics.

Our main finding is that discrete kink ratchets in the absence of
PNp, at least for relatively small amplitude of driving force, is
very much different from the conventional discrete kink ratchets
experiencing PNp (compare the results plotted in Fig. \ref{fig5}
and Fig. \ref{fig6}).

Particularly, the acceleration of the PNp-free kink due to AC
driving practically does {\em not} depend on $h$ in the
non-resonance range of the driving frequency $\Omega$ (see Fig.
\ref{fig8}). Indeed, in the range of $0.1 < \Omega <1.2$ we have
practically same $a(\Omega)$ dependence for relatively weak
($h=0.3$), moderate ($h=0.6$), and strong ($h=0.9$) discreteness.
Is it is well-known that typically, physical properties of a
discrete system are extremely sensitive to the discreteness
parameter $h$. This unusual result can be explained by the fact
that for the static kinks in DKGM2 PNp is precisely equal to zero.

Influence of $h$ on $a(\Omega)$ appears only through the influence
of the kink's internal modes whose frequencies are $h$-dependent
[see Fig. \ref{fig3}(b)].

We also confirm earlier findings
\cite{BraunKivshar:2004,Modes1,Cont5,Cont6,Discrete1} that the
efficiency of ratchets, measured in our undamped case by the
acceleration of kink, considerably increases when the driving
force frequency approaches kink's internal mode frequency (see
Fig. \ref{fig7} and Fig. \ref{fig8}). We also observed that $a
\sim A^2$, where $a$ is kink's acceleration and $A$ is the driving
force amplitude. This is similar to the scaling rule $\left\langle
v \right\rangle \sim A^2$ reported earlier for the averaged
soliton velocity, $\left\langle v \right\rangle$,
\cite{BraunKivshar:2004}.

Main reason for the striking difference in the kink ratchet
dynamics observed for PNp-free and ordinary kinks lies in the fact
that the static PNp-free kinks are not trapped by the lattice and
they possess the zero-frequency translational Goldstone mode for
any degree of discreteness [see Fig. \ref{fig3}(b)].

Many interesting problems are left out of the scope of the present
work, such as influence of damping, the effect of asymmetric,
e.g., biharmonic AC driving, stochastic driving, etc. We plan to
address these issues in forthcoming publications.

\section*{Acknowledgements}
SVD wishes to thank the warm hospitality of the Institute of
Physics in Bhubaneswar, India. This work was supported by the
RFBR-DST Indo-Russian grant 08-02-91316-Ind-a and by the RFBR
grant 09-08-00695-a.



\end{document}